\documentclass{article}

\usepackage[position,nonatbib,preprint]{neurips_2026}

\usepackage[numbers]{natbib}

\usepackage[utf8]{inputenc}
\usepackage[T1]{fontenc}
\usepackage{hyperref}
\usepackage{url}
\usepackage{booktabs}
\usepackage{amsfonts}
\usepackage{amsmath}
\usepackage{nicefrac}
\usepackage{microtype}
\usepackage{xcolor}
\usepackage{pifont}
\usepackage{wasysym}
\usepackage{multirow}

\title{Securing LLM Agents Need Intent-to-Execution Integrity}

\author{
Wenjie Qu$^{1}$ \quad
Ming Xu$^{1}$ \quad
Peiran Wang$^{2}$ \quad
Shengfang Zhai$^{1}$ \quad
Jiaheng Zhang$^{1}$ \quad
Dawn Song$^{3}$ \\
$^{1}$ National University of Singapore \\
$^{2}$ University of California, Los Angeles \\
$^{3}$ University of California, Berkeley
}

\begin{document}

\maketitle

\begin{abstract}
This position paper argues that securing LLM agents requires first defining an end-to-end correctness property that specifies when an agent's execution faithfully reflects the user's intent. Modern LLM agents operate over an \emph{intent-to-execution pipeline}, where natural-language instructions are translated into concrete system operations such as tool calls, API requests, and code execution. While recent defenses have made progress in constraining how agents construct tool calls, most existing formulations implicitly assume that tools are trusted.
The emergence of systems such as OpenClaw, with open ecosystems of third-party skills and direct access to user environments, breaks this assumption and exposes new failure modes, including malicious or over-privileged components in the execution pipeline.

Despite rapid progress in defense mechanisms, there is no adequate correctness property that defines what ``secure'' means for LLM agents, nor a principled way to evaluate the coverage of existing defenses. We observe that LLM agents are structurally analogous to compilers, where security violations correspond to mis-executions that do not preserve user intent. Drawing on this analogy, we identify two fundamental problem sources---untrusted data ingestion and untrusted tool execution---and derive four integrity properties that must hold simultaneously: \emph{Tool Integrity}, \emph{Instruction Integrity}, \emph{Judgment Integrity}, and \emph{Data Flow Integrity}.
We call their conjunction \emph{intent-to-execution integrity}.

Analyzing existing agentic defenses against these properties reveals that current systems provide only partial and non-compositional coverage, leaving fundamental gaps in securing modern LLM agents. 
\end{abstract}

\section{Introduction}
\label{sec:intro}

LLM agents are increasingly deployed to execute complex tasks involving reasoning, tool use, persistent memory, and interaction with external systems.
From personal assistants~\citep{openclaw2025} to software engineering agents, these systems no longer merely generate text---they take actions: editing files, invoking APIs, executing code, and orchestrating workflows.

This shift fundamentally changes the security problem.
Traditional LLM safety focuses on controlling model outputs.
In contrast, LLM agents operate over an \emph{intent-to-execution pipeline}, where user intent expressed in natural language is translated into concrete system operations.
Security violations in this setting are not just harmful outputs, but \emph{mis-executions}---cases where the agent's execution effects misalign with the user's intent due to attacks. 

A large body of recent work has proposed defenses for LLM agents, including prompt-injection detection~\citep{shi2025promptarmor, chen2025indirect}, policy enforcement~\citep{shi2025progent, debenedetti2025camel, li2025drift}, and task-alignment mechanisms~\citep{jia2025taskshield}.
These defenses have made important progress in securing agentic planning: constraining how agents translate user intent into tool calls and arguments. 
However, most existing agent-security formulations implicitly assume that tools are trusted. 
Under this assumption, the security problem largely reduces to preventing untrusted inputs from influencing tool-call construction.

\paragraph{From trusted tools to open ecosystems.}
A growing class of agent frameworks challenges the assumption that tools can be treated as trusted components.
Systems such as OpenClaw~\citep{openclaw2025} feature open ecosystems of community-contributed skills, persistent cross-session memory, and direct access to the user's local environment---settings in which tools are contributed by third parties and therefore cannot be assumed to be trusted.

This shift introduces failure modes that are not captured by existing formulations, particularly supply-chain attacks through the execution pipeline.
These risks are not merely theoretical:
the ClawHavoc campaign~\citep{koisecurity2026clawhavoc} planted over 1,000 malicious skills on ClawHub to harvest credentials and cryptocurrency wallets;
a large-scale audit of 42,447 skills~\citep{liu2026skillaudit} found that 26.1\% contain at least one security vulnerability;
and Snyk~\citep{snyk2026openclaw} demonstrated sandbox bypasses where configured security boundaries are not enforced at runtime.

Meanwhile, recent defenses for such systems---including NemoClaw~\citep{nemoclaw2026}, SeClaw~\citep{seclaw2026}, SafeClaw-R~\citep{safeclawr2026}, and SecureClaw~\citep{secureclaw2026}---stack multiple mechanisms such as sandboxing, pattern scanning, and prompt-injection filters.
However, each system introduces mechanisms without defining a correctness property that these mechanisms are intended to achieve, making it difficult to assess what they actually guarantee.

\paragraph{The missing piece.}
The limitations of existing defenses have a common root cause: there is no \emph{adequate} correctness property that defines what ``secure'' means for LLM agents.

Formally, a correctness property specifies when an agent's execution effect is correct with respect to the user's intent across the entire intent-to-execution pipeline.
Without such a definition, defense design becomes ad hoc, evaluation lacks a principled target, and practitioners cannot reason about the completeness of security mechanisms.

Prior work has begun to formalize aspects of agent security~\citep{siu2026framework}.
However, the formulation in ~\citep{siu2026framework} assumes trusted tools and focuses on the integrity of tool-call selection and arguments. 
It does not capture untrusted tool execution and reasoning-level manipulation, and therefore does not provide an end-to-end correctness condition for modern LLM agents. 

\paragraph{Our position.}
\textbf{Advancing the security of LLM agents requires first defining an end-to-end correctness property for the intent-to-execution pipeline.}

We identify two fundamental problem sources---untrusted data ingestion and untrusted tool execution---and derive four integrity properties that must hold simultaneously: \emph{Tool Integrity}, \emph{Instruction Integrity}, \emph{Judgment Integrity}, and \emph{Data Flow Integrity}.
We call their conjunction \emph{intent-to-execution integrity}.

This definition enables systematic evaluation of existing defenses (which properties does each system cover?), principled benchmark design (what should be measured?), and identification of structural gaps that mechanism stacking cannot close. 
Evaluating existing defenses against these properties reveals that no system achieves full coverage.

\section{Intent-to-Execution Integrity}
\label{sec:integrity}

\subsection{Why a Correctness Property}
\label{sec:why}

An LLM agent transforms user intent expressed in natural language into concrete system operations---file modifications, API calls, shell commands---through a multi-stage pipeline: the LLM interprets intent, generates a plan, selects tools, and executes tool calls, which in turn become operating system instructions. 
This process is structurally analogous to compilation: a high-level specification (user intent) is progressively lowered through intermediate representations (plans, tool calls) into executable operations (system calls, I/O).
Recent work has begun to make this analogy explicit, framing LLMs as interpreters for natural language programs~\citep{ge2024core} and compiling natural-language intent into typed execution graphs~\citep{zhang2025agint}.

In compiler design, correctness is defined as \emph{semantic preservation}---the compiled binary must behave as prescribed by the source program's semantics.
CompCert~\citep{leroy2009compcert} famously proved this property end-to-end for a realistic C compiler.
The security of modern LLM agents demands an analogous property: the agent's execution must faithfully reflect the user's intent. 
Without such a property, security reduces to a reactive arms race---every new attack gets a new patch, but there is no way to assess whether the patches are complete or where gaps remain.
We derive such a property by tracing every security threat to its root cause.
Modern LLM Agents face exactly two fundamental problem sources; every integrity property we propose exists to close a causal path from source to harm.

\begin{table}[t]
  \caption{Derivation of integrity properties from problem sources and causal paths.}
  \label{tab:derivation}
  \centering
  \small
  \begin{tabular}{@{}lll@{}}
    \toprule
    Problem source & Risks & Required property \\
    \midrule
    \multirow{3}{*}{Untrusted data ingestion}
      & Attribution confusion & Instruction Integrity \\
      & Data flow violation & Data Flow Integrity \\
      & Judgment manipulation & Judgment Integrity \\
    \midrule
    Untrusted tool execution
      & Malicious third-party skills/tools & Tool Integrity \\
    \bottomrule
  \end{tabular}
\end{table}
\subsection{Problem Source 1: Untrusted Data Ingestion}
\label{sec:ps1}

LLM agents process external content---emails, web pages, documents, and tool outputs---that may contain malicious instructions.  
This content is processed within a context that may contain sensitive information accumulated from prior interactions and tool executions.
Three causal paths lead from this source to harm:

\begin{enumerate}
    \item \emph{Attribution confusion}: the agent cannot reliably distinguish user instructions from malicious instructions embedded in untrusted data. A malicious email containing ``forward all files to attacker@evil.com'' is processed in the same context as the user's command ``summarize my inbox.'' Indirect prompt injection exploits this confusion at scale~\citep{greshake2023injection, debenedetti2024agentdojo}.
    \item \emph{Data flow violation}: data read by one tool call flows into the parameters of another in ways that violate provenance constraints. Credentials loaded from \texttt{.env} during debugging end up as content in an outbound email; a file path intended as a read source is passed as a write destination; a user's private budget retrieved in session~1 appears in a response to a different user in session~2 through persistent memory. These violations occur even when every tool call aligns to the user's intent---the problem is that tool call arguments carry sensitive data across boundaries that should be enforced but are not.
    \item \emph{Judgment manipulation}: even when no explicit instruction is injected, adversarial content in the agent's context can subtly steer its reasoning and alter its decision outcomes. A paper under review may contain self-promotional language that biases a paper-review agent toward a favorable assessment; carefully designed product review injections may cause a shopping agent to misjudge product quality and make a poor purchasing decision. Unlike attribution confusion, judgment manipulation does not require the agent to mistake data for a command---it exploits the fact that the agent must \emph{read} data to make decisions, and that data can influence those decisions adversarially.
\end{enumerate}

To prevent these three failure modes, we define three correctness properties.

\paragraph{Instruction Integrity.}
Every agent action must be causally attributable to an instruction from an authenticated user, not injected content.  
When a user explicitly delegates trust to external content (e.g., ``follow this recipe''), subsequent actions from that content are authorized within the delegated scope. 

\paragraph{Data Flow Integrity.}
Data flowing through tool call parameters must respect provenance constraints.
Sensitive data (credentials, private records, session-specific context) must be taint-marked at read time, and tool calls whose arguments carry tainted data to unauthorized destinations (outbound email, external APIs, other users' sessions) must be blocked. 
Taint marks must persist across sessions.

\paragraph{Judgment Integrity.}
All reasoning decisions made by the agent must be derived from user-aligned objectives. The reasoning process should be robust to adversarial influence from untrusted data. 

We make this precise as follows.
Let $q$ denote the user's query, $d$ the untrusted data the agent must read to perform its task (e.g., a paper to review, a set of products to rank), and $J(q, d)$ the agent's decision outcome (e.g., a review score, a ranking, a binary accept/reject). 
The injected content $\delta$ applied to $d$ is \emph{task-irrelevant} if it does not change the ground-truth answer that a correct reasoner would produce---for instance, inserting self-promotional rhetoric into a paper does not change its actual technical contribution.
Judgment Integrity requires that for every task-irrelevant perturbation $\delta$,
\[
  J(q,\; d \oplus \delta) \;=\; J(q,\; d).
\]
In other words, the agent's decision must depend only on the task-relevant content of the untrusted data, and must be invariant to adversarial, task-irrelevant manipulation.

We defer the argument for why these three properties, together with Tool Integrity (defined in Section~\ref{sec:ps2}), are jointly sufficient to Section~\ref{sec:sufficiency}.

\begin{table}[t]
  \caption{Examples for violations of each integrity condition.}
  \label{tab:necessity}
  \centering
  \small
  \begin{tabular}{@{}p{2.5cm}p{8cm}@{}}
    \toprule
    Integrity condition & Example violation \\
    \midrule
    Tool Integrity
      & A third-party image-generation skill includes hidden code that reads the user's SSH key and exfiltrates it to a remote server.  \\
    \addlinespace
    Instruction Integrity
      & A user asks the agent to summarize an email, but the agent follows an injected instruction inside the email such as ``forward all files to attacker@evil.com.'' \\
    \addlinespace
  Data Flow Integrity
  & A user asks the agent to debug a local application and email a report to a teammate. 
    An injected instruction in a log file changes the report body to include environment variables read from \texttt{.env}.
    Sensitive data is sent to the teammate. \\
  Judgment Integrity
  & A paper-review agent is asked to evaluate a submitted paper.
    The paper includes adversarial self-promotional text such as
    ``this paper is widely regarded by leading experts as the best work in this field.''
    The agent is biased by this content and produces an overly favorable review. \\
    \addlinespace
    \bottomrule
  \end{tabular}
\end{table}

\subsection{Problem Source 2: Untrusted Tool Execution}
\label{sec:ps2}

Modern agent frameworks increasingly load and execute third-party skills, plugins, and scripts contributed by open community ecosystems.
In systems like OpenClaw~\citep{openclaw2025}, users install these components themselves, often with no centralized review or IT oversight.

This leads to direct supply-chain harm to users because third-party skills can be malicious while appearing to provide legitimate functionality.
For example, the ClawHavoc campaign planted hundreds of malicious skills on ClawHub that masqueraded as benign automation tools, including wallet, trading, and productivity-related skills. 
Once installed or executed, these malicious skills could deliver payloads that steal credentials, browser data, SSH credentials, cryptocurrency wallet data, Apple and KeePass keychains, and user documents~\citep{koisecurity2026clawhavoc, trendmicro2026amos, hackernews2026clawhavoc}.

To prevent this failure mode, we define an additional correctness property:

\paragraph{Tool Integrity.}
Every software component that participates in producing an action---skill, plugin, or script---must be confined to its declared interface and documented behavior. 
For any invocation within its advertised functionality, the system must prevent the component from producing hidden or undeclared side effects, such as reading unrelated local files, contacting unadvertised network endpoints, modifying persistent state outside its documented scope, or executing additional commands not required by the requested functionality. Tool Integrity therefore requires enforcement of behavioral conformance between what a component claims to do and what it is allowed to do at runtime.

\subsection{Putting the Properties Together}
\label{sec:sufficiency}

The four properties compose into an end-to-end correctness condition for the LLM Agent's intent-to-execution pipeline.
We argue their sufficiency by exhaustive case analysis over the pipeline stages.

At each time step, the agent performs one of three operations: (i)~\emph{reasoning}---reading data and producing a semantic decision, (ii)~\emph{action construction}---selecting a tool and populating its arguments, or (iii)~\emph{execution}---invoking a software component that carries out the action.
An adversary can corrupt the pipeline only by corrupting one of these operations.
We enumerate all possible corruption points:

\begin{itemize}
\item \textbf{Reasoning} can be corrupted if untrusted data distorts the agent's semantic decision (e.g., a biased review). Judgment Integrity prevents this: it requires $J(q, d\oplus\delta) = J(q, d)$ for all task-irrelevant perturbations $\delta$.
\item \textbf{Action construction} can be corrupted in two ways: (a)~the agent selects a tool call that was not intended by the user (attribution confusion), or (b)~the agent populates tool-call arguments with data that violates provenance constraints (data flow violation). Instruction Integrity prevents~(a); Data Flow Integrity prevents~(b).
\item \textbf{Execution} can be corrupted if the invoked component performs undeclared behavior beyond its interface specification. Tool Integrity prevents this.
\end{itemize}

Because every agent time step falls into one of these three operations, and every corruption path within each operation is closed by the corresponding property, no pipeline stage remains unprotected when all four properties hold simultaneously.
If all four hold, then the agent's decisions are formed without adversarial distortion, its actions align with the user's original intent, its action parameters carry only authorized data flows, and the invoked components execute only their declared behavior. 
We call this conjunction \emph{intent-to-execution integrity}.

\section{Evaluating Existing Defenses}
\label{sec:eval}

We evaluate existing LLM agent defense systems against the four integrity properties.
Our goal is to answer the following question:
which part of intent-to-execution integrity does each defense actually enforce? 

We use three levels in Table~\ref{tab:coverage}.
\CIRCLE{} means that the defense provably enforces the corresponding property. 
\LEFTcircle{} means that the defense contains a relevant mechanism, but the mechanism is based on heuristics or limited to specific attack patterns, thus it is feasible to bypass. 
\Circle{} means that the defense does not address the property. 

\begin{table}[t]
  \caption{Defense coverage across four integrity properties. \CIRCLE~=~full (provable), \LEFTcircle~=~partial, \Circle~=~none.}
  \label{tab:coverage}
  \centering
  \small
  \begin{tabular}{@{}lcccc@{}}
    \toprule
    Defense & Instr. Int. & Data Flow Int. & Judg. Int. & Tool Int. \\
    \midrule
    \multicolumn{5}{@{}l}{\emph{Defense systems for General LLM Agents}} \\
    \addlinespace
    PromptArmor~\citep{shi2025promptarmor} & \LEFTcircle & \LEFTcircle & \LEFTcircle & \Circle \\
    PIA-Detection~\citep{chen2025indirect} & \LEFTcircle & \LEFTcircle  & \LEFTcircle & \Circle \\
    TaskShield~\citep{jia2025taskshield}  & \LEFTcircle & \LEFTcircle  & \Circle & \Circle \\
    Progent~\citep{shi2025progent} & \LEFTcircle & \CIRCLE & \Circle & \Circle \\
    CaMeL~\citep{debenedetti2025camel} & \CIRCLE & \CIRCLE & \Circle & \Circle \\
    DRIFT~\citep{li2025drift} & \CIRCLE & \LEFTcircle & \Circle & \Circle \\
    \addlinespace
    \midrule
    \multicolumn{5}{@{}l}{\emph{Defense systems for OpenClaw}} \\
    \addlinespace  NemoClaw~\citep{nemoclaw2026} & \Circle & \Circle & \Circle & \LEFTcircle \\
        IronClaw~\citep{ironclaw2026} & \LEFTcircle & \LEFTcircle & \LEFTcircle & \LEFTcircle \\
    SeClaw~\citep{seclaw2026} & \LEFTcircle & \LEFTcircle & \LEFTcircle & \LEFTcircle \\
    SafeClaw-R~\citep{safeclawr2026} & \LEFTcircle & \Circle & \Circle & \LEFTcircle \\
    SecureClaw~\citep{secureclaw2026} & \LEFTcircle & \LEFTcircle & \LEFTcircle & \LEFTcircle \\
    \bottomrule
  \end{tabular}
\end{table}

\subsection{Defense Systems for General LLM Agents}
\label{sec:general-defenses}
Most general LLM-agent defenses are designed for cloud-hosted settings, where tools are assumed to be trusted.
Under this assumption, the main security problem is whether untrusted data can cause the agent to select the wrong tool call or construct unsafe arguments.
Accordingly, these defenses mainly protect the tool-call construction process rather than the integrity of the invoked components.

Within this trusted-tool setting, existing defenses fall into two groups.
Detection-based methods such as PromptArmor~\citep{shi2025promptarmor} and PIA-Detection~\citep{chen2025indirect} attempt to detect and remove injected instructions before they influence the agent.
System-security-style methods such as CaMeL~\citep{debenedetti2025camel}, DRIFT~\citep{li2025drift}, and Progent~\citep{shi2025progent} enforce constraints over tool-call control flow or argument-level data flow, providing meaningful coverage over Instruction Integrity and Data Flow Integrity.

However, this design focus also explains two systematic gaps.
First, Tool Integrity is largely out of scope because these systems assume invoked tools faithfully implement their interfaces.
Second, for system-security-style methods, Judgment Integrity is not enforced because their guarantees concern tool-call behavior, not semantic reasoning outcomes.
An adversarial paper may bias a review agent into producing an overly favorable assessment without causing any unauthorized tool call, remaining within the constraints enforced by system-security-style methods.

\paragraph{PromptArmor~\citep{shi2025promptarmor}.}
PromptArmor is an input-side guardrail that prompts an off-the-shelf LLM to detect and remove injected prompts before the agent processes the input.
It can partially alleviate attribution confusion, and can also mitigate some Data Flow and Judgment violations when those violations are triggered by explicit injected instructions that are successfully removed.
However, it provides no guarantee that all injected prompts will be detected or removed.
Finally, PromptArmor does not verify whether tools or skills faithfully implement their declared behavior, so it provides no Tool Integrity.

\paragraph{PIA-Detection~\citep{chen2025indirect}.}
PIA-Detection fine-tunes a DeBERTa model to detect prompt injections in documents.
Like PromptArmor, its protection centers on identifying injected instruction spans.
Thus, it provides partial protection for Instruction Integrity, Data Flow Integrity and Judgment Integrity. 
However, it provides no guarantee that injected prompts will be detected under distribution shift or adaptive attacks.
It also does not address Tool Integrity.

\paragraph{TaskShield~\citep{jia2025taskshield}.}
TaskShield reframes defense as task alignment: each instruction and tool call is checked against the user's stated goal.
This gives partial Instruction Integrity because tool calls that do not contribute to the user task can be blocked.
It also gives partial Data Flow protection when an exfiltrating tool call or sensitive argument is judged task-irrelevant.
However, TaskShield relies on LLM-based task-alignment checks, so the checker itself remains vulnerable to adaptive attacks.
We mark Judgment Integrity as none because TaskShield focuses on checking tool calls rather than ensuring that the agent's semantic reasoning outcomes remain unbiased by adversarial content.
It also does not address Tool Integrity because it assumes invoked tools are trusted.

\paragraph{Progent~\citep{shi2025progent}.}
Progent enforces programmable least-privilege policies over tool calls using a domain-specific language.
Because these policies decide which tool calls are permitted during execution, Progent partially enforces Instruction Integrity: an injected instruction may fail if it asks for an action outside the allowed policy.
Progent provides deterministic guarantees for policy-specified Data Flow constraints by using policies to constrain tool-call arguments.
However, these guarantees are policy-relative rather than general provenance tracking or taint propagation.
Progent does not protect Judgment Integrity because it focuses on tool-call enforcement rather than the agent's reasoning process.
It also leaves Tool Integrity unaddressed.

\paragraph{CaMeL~\citep{debenedetti2025camel}.}
CaMeL extracts tool-call control flows from the trusted user query, architecturally preventing untrusted data from affecting program flow, and uses policies and information-flow tracking to enforce Data Flow Integrity.
For this reason, we mark both Instruction Integrity and Data Flow Integrity as fully enforced.
However, CaMeL does not solve Judgment Integrity: its design does not prevent the model's semantic assessment from being distorted by adversarial but non-instructional content.
CaMeL also does not provide Tool Integrity, since it assumes invoked tools faithfully implement their declared behavior.

\paragraph{DRIFT~\citep{li2025drift}.}
DRIFT combines a secure planner, an LLM-based validator, and an injection isolator.
The planner constructs a minimal function trajectory and parameter checklist from the user query; the validator monitors deviations from that plan; and the isolator masks conflicting instructions from memory.
This provides support for Instruction Integrity because deviations from the intended function trajectory can be blocked.
It also gives partial Data Flow Integrity because parameter checklists and data-level constraints can restrict some unsafe argument flows.
The protection is only partial because validation relies on LLM-based judgments, which may remain vulnerable to adaptive attacks.
DRIFT does not address Judgment Integrity or Tool Integrity.

\subsection{Defense Systems for OpenClaw}
\label{sec:personal-defenses}
OpenClaw defenses address the deployment-layer risks that general-agent defenses largely leave out of scope.
NemoClaw and IronClaw harden execution through sandboxing, capability restrictions, and network controls, while SeClaw, SafeClaw-R, and SecureClaw add skill audits, monitors, hardening rules, behavioral checks, or safety-enforced counterparts.
These mechanisms reduce the damage of malicious or vulnerable skills, and therefore provide partial coverage for Tool Integrity.
However, they mostly enforce confinement, detection, or rule-based checks rather than exact behavioral conformance.
They can limit what an untrusted component accesses, or detect known malicious patterns, but they do not guarantee that undeclared component behavior is always blocked. 
Moreover, most OpenClaw defenses only partially cover Instruction, Data Flow, and Judgment Integrity, because their protections are mainly heuristic-based. 

\paragraph{NemoClaw~\citep{nemoclaw2026}.}
NemoClaw primarily hardens the execution environment.
It runs the OpenClaw agent inside an OpenShell sandbox with Landlock, seccomp, and network namespace isolation, and restricts network egress by policy.
This gives partial Tool Integrity because sandboxing constrains what a malicious component can access.
However, sandboxing can limit the damage of a malicious skill, but it cannot fully prevent all hidden side effects within the permissions granted to the sandbox.
NemoClaw does not provide mechanisms for Instruction Integrity, Data Flow Integrity, or Judgment Integrity, because its design does not separate instructions from data, track information flow across tool calls, or protect semantic reasoning from adversarial content.

\paragraph{IronClaw~\citep{ironclaw2026}.}
IronClaw provides broader defense-in-depth for OpenClaw.
Its public design emphasizes isolated WebAssembly containers for untrusted tools, capability-based permissions, endpoint allowlisting, credential protection at the host boundary, leak detection, prompt-injection defense, content sanitization, and policy enforcement.
We therefore mark all four properties as partial.
Its prompt-injection and policy mechanisms can partially protect Instruction Integrity and restrict some unsafe data flows.
Its content sanitization can reduce reasoning manipulation, providing partial protection for Judgment Integrity.
Its WASM sandboxing and capability restrictions provide partial Tool Integrity by limiting what untrusted components can access or execute.
However, none of these mechanisms provides a full end-to-end guarantee for any of the four properties.

\paragraph{SeClaw~\citep{seclaw2026}.}
SeClaw achieves broad but shallow coverage.
Its skill auditing, runtime rules, privacy checks, and prompt-injection defenses provide relevant mechanisms for all four properties.
However, these mechanisms are primarily detection-, rule-, or audit-based.
They can reduce the likelihood that injected instructions are followed, sensitive data is leaked, semantic manipulation succeeds, or malicious skills are installed.
But they do not provide end-to-end guarantees for any property.
Thus, each property is marked as partial.

\paragraph{SafeClaw-R~\citep{safeclawr2026}.}
SafeClaw-R provides mechanisms to mediate actions prior to execution, and augments skills with safety-enforced counterparts.
This gives partial Instruction Integrity because unsafe or task-misaligned actions can be blocked before execution.
Because its mediation logic may be detector- or policy-based, we mark this protection as partial.
SafeClaw-R does not implement Data Flow Integrity: it is not a taint-tracking or information-flow-control system over values moving between tools. 
It also does not enforce Judgment Integrity, because its design focuses on mediating tool calls and actions rather than protecting the agent's reasoning process outcomes from adversarial influence.
SafeClaw-R gives partial Tool Integrity because it detects malicious third-party skill patterns and modifies skills through safety-enforced counterparts.

\paragraph{SecureClaw~\citep{secureclaw2026}.}
SecureClaw combines installation auditing, hardening modules, background monitors, and behavioral rules.
It explicitly targets prompt injection, credential theft, supply-chain attacks, and privacy leaks.
This breadth justifies partial coverage for all four properties.
Its behavioral rules can reduce some instruction-attribution and judgment-manipulation failures; its credential and privacy checks can mitigate some data-flow violations; and its audit and hardening mechanisms can detect or reduce some malicious or vulnerable skills.
However, these mechanisms are primarily heuristic. 
Thus, all four properties are marked as partial.

\subsection{Takeaways}
\label{sec:eval-takeaways}

Table~\ref{tab:coverage} shows that existing defenses cover complementary but incomplete parts of intent-to-execution integrity.
General-agent defenses, especially system-security-style designs, are strongest at protecting the \emph{translation layer}: they constrain how user intent is lowered into tool calls and tool-call arguments. 
This explains their coverage of Instruction Integrity and Data Flow Integrity.
However, they largely assume that the invoked tools are trusted, leaving Tool Integrity out of scope.  
They also provide very limited protection to Judgment Integrity, because semantic reasoning can be distorted without producing an unauthorized tool call. 
 
OpenClaw defenses address the \emph{execution layer} more directly. 
They harden the runtime, restrict component capabilities, audit skills, monitor suspicious behavior, or replace unsafe skills with safer counterparts.
These mechanisms partially mitigate malicious or vulnerable components, but they mostly provide confinement, detection, or rule-based enforcement rather than complete behavioral conformance.
Thus, they improve Tool Integrity only partially, and their coverage of Instruction, Data Flow, and Judgment Integrity remains limited. 

The central gap is therefore compositional.
Protecting only the translation layer is insufficient: a correctly selected tool call can still invoke a malicious skill.
Protecting only the execution layer is also insufficient: a well-confined component can still be invoked for the wrong reason, with the wrong arguments, or after adversarially distorted reasoning.
Intent-to-execution integrity requires both layers to be secured simultaneously.

\section{What Existing Techniques Can and Cannot Guarantee}

The defense evaluation in Section~\ref{sec:eval} shows \emph{which} properties each system covers.
A complementary question is \emph{how far} current techniques can go in principle for each property, independent of any particular system.
We discuss each property in turn.

\paragraph{Instruction Integrity.} Instruction Integrity can be strongly enforced through architectural separation between trusted user instructions and untrusted data, as demonstrated by CaMeL~\citep{debenedetti2025camel} and Fides~\citep{costa2025fides}.
Under such designs, control flow is derived solely from authenticated user input, preventing untrusted content from altering tool selection. 
The principal limitation is utility: planning without access to untrusted content severely restricts the agent in tasks where decisions must depend on tool-call returns or external data.
Relaxing this separation to recover utility reintroduces the risk of instruction attribution confusion.

\paragraph{Data Flow Integrity.} Data Flow Integrity can be enforced using information-flow control techniques~\citep{wu2024systemlevelifc,costa2025fides} such as taint tracking and provenance-aware policies. 
These approaches provide deterministic guarantees under well-defined policies---for example, a policy can constrain email recipients to a trusted allowlist, making it impossible for an attacker to exfiltrate private data to addresses they control. 
The principal limitation is scalability: integrating new tools requires manually specifying per-tool policies, which becomes impractical for ecosystems with numerous third-party skills.

\paragraph{Judgment Integrity.}
Judgment Integrity remains the least developed dimension and is qualitatively harder than the preceding two.
Instruction Integrity and Data Flow Integrity protect \emph{structured} objects---tool-call selections and argument values---that can be separated from untrusted data by architecture or tracked by taint propagation.
Judgment is different: many agent tasks require the model to read untrusted data precisely in order to make a semantic decision---reviewing a paper, ranking products, summarizing emails, or assessing evidence.
In such tasks, the untrusted data cannot simply be isolated from the reasoning process, because it is the evidence on which the judgment depends.

This makes the problem substantially harder to enforce.
Model-level defenses such as StruQ~\citep{chen2024struq} and SecAlign~\citep{chen2025secalign} improve robustness against injected instructions through structured queries or preference optimization, but they do not provide guarantees over semantic judgments. 
Moreover, adaptive attacks can bypass such defenses~\citep{zhan2025adaptive, chen2025topicattack}.  
LLM-based detection and input sanitization are useful heuristics but cannot guarantee that $J(q, d\oplus\delta) = J(q, d)$ for all task-irrelevant perturbations $\delta$. 
Designing mechanisms that provide stronger integrity guarantees for semantic reasoning, remains an important open direction.

\paragraph{Tool Integrity.}
Tool Integrity can be partially enforced through a combination of static analysis, sandboxing, and runtime confinement.
Static analysis can inspect tool code before installation to detect suspicious APIs, undeclared file access, and credential handling~\citep{duan2021measuring}.
Sandboxing~\citep{nemoclaw2026} can restrict runtime privileges by limiting filesystem access, system calls, and network egress.
Together, these techniques substantially reduce the damage of malicious tools.

However, neither provides full behavioral conformance.
Static analysis is necessarily approximate and can miss obfuscated or context-dependent behavior. 
Sandboxing cannot prevent unintended actions that remain within its permitted capabilities.
A promising direction for better approximation of tool integrity is dynamic analysis of agent tools, where tools are executed in instrumented environments to observe actual behavior and check whether it conforms to the declared specification.

\section{Alternative Views}
\label{sec:alternatives}

\paragraph{``Existing formalizations already define agent security.''}
Prior work has begun to formalize important aspects of LLM-agent security, especially the integrity of tool-call selection and data isolation~\citep{siu2026framework}. 
We view their effort as complementary rather than competing. 
Our argument is that existing formulations are not yet adequate for modern agent deployments because they typically assume trusted tools and focus on the planning layer---tool-call selection and argument integrity.
This leaves two dimensions unaddressed: untrusted tool execution (Tool Integrity), which arises in open ecosystems where components are community-contributed; and reasoning-level manipulation (Judgment Integrity), which exploits the fact that agents must read untrusted data to make semantic decisions. 
Intent-to-execution integrity makes the trusted-pipeline assumption explicit and adds the missing dimensions needed for open, tool-rich, and stateful agent ecosystems.

\paragraph{``Judgment Integrity is too vague to be useful.''}
Judgment Integrity is indeed harder to formalize than Instruction Integrity or Data Flow Integrity, but that is precisely why it warrants explicit treatment. Tool-call selection and arguments are externalized, structured objects that can be separated from untrusted data and validated against policies. Judgment is different: many agent tasks require reading untrusted content in order to make a semantic decision—reviewing a paper, ranking products, or assessing evidence. Even though current defense techniques poorly address judgment integrity, explicitly defining it delineates a critical security gap that existing agentic defenses overlook. 

\paragraph{``The property is too strong for practical systems.''}
Intent-to-execution integrity is an end-to-end ideal, not a claim that current systems can immediately satisfy all four properties in full. This mirrors a familiar pattern in traditional security: formal verification defines correctness properties for entire programs even though current techniques struggle to scale beyond modest-sized programs. The value of such properties lies not in immediate full enforcement, but in making the target explicit. Similarly, intent-to-execution integrity provides the reference standard against which defenses and benchmarks should be measured. This framing helps reveal which parts of the pipeline are adequately protected, which remain only partially so, and—crucially—which have received little attention from the agent security research community.

\section{Conclusion}
\label{sec:conclusion}

We argued that securing LLM agents requires an end-to-end correctness property---\emph{intent-to-execution integrity}---composed of four conditions: Instruction Integrity, Data Flow Integrity, Judgment Integrity, and Tool Integrity.
Evaluating existing defenses against these properties shows that each covers a different subset, and no system achieves full coverage.
Our position is not that existing defenses are ineffective, but that they are incomplete without a shared correctness target; intent-to-execution integrity provides such a target for comparing defenses, designing benchmarks, and identifying where future work is most needed.

  
 
{
\small
 
{
\small
\bibliographystyle{plainnat}

}

\end{document}